\documentclass[aps,prb,nobibnotes,twocolumn,superscriptaddress,bibliography]{revtex4-2}

\pdfoutput=1
\usepackage{amsfonts}
\usepackage{mathrsfs}
\usepackage{amsmath}
\usepackage{color}
\usepackage{graphicx}
\graphicspath{ {./pictures/version_9/} }
\usepackage{bm}
\usepackage{amssymb}
\usepackage{xspace}
\usepackage{epstopdf}
\usepackage{dcolumn}
\usepackage{longtable}
\usepackage{multirow}
\usepackage{float}
\usepackage{comment}

\providecommand{\tabularnewline}{\\}

\usepackage[colorlinks=true, letterpaper=true, pdfstartview=FitV, linkcolor=blue, citecolor=blue, urlcolor=blue]{hyperref}

\newcommand{\Rom}[1]{\uppercase\expandafter{\romannumeral#1}}

\makeatother

\begin{document}

\title{Planar Hall effect in topological Weyl and nodal line semimetals}%

\author{Lei Li}
\affiliation{Centre for Quantum Physics, Key Laboratory of Advanced Optoelectronic Quantum Architecture and Measurement (MOE), School of Physics, Beijing Institute of Technology, Beijing, 100081, China}
\affiliation{Beijing Key Lab of Nanophotonics \& Ultrafine Optoelectronic Systems, School of Physics, Beijing Institute of Technology, Beijing, 100081, China}

\author{Jin Cao}
\affiliation{Centre for Quantum Physics, Key Laboratory of Advanced Optoelectronic Quantum Architecture and Measurement (MOE), School of Physics, Beijing Institute of Technology, Beijing, 100081, China}
\affiliation{Beijing Key Lab of Nanophotonics \& Ultrafine Optoelectronic Systems, School of Physics, Beijing Institute of Technology, Beijing, 100081, China}

\author{Chaoxi Cui}
\affiliation{Centre for Quantum Physics, Key Laboratory of Advanced Optoelectronic Quantum Architecture and Measurement (MOE), School of Physics, Beijing Institute of Technology, Beijing, 100081, China}
\affiliation{Beijing Key Lab of Nanophotonics \& Ultrafine Optoelectronic Systems, School of Physics, Beijing Institute of Technology, Beijing, 100081, China}

\author{Zhi-Ming Yu}
\email{zhiming\_yu@bit.edu.cn}
\affiliation{Centre for Quantum Physics, Key Laboratory of Advanced Optoelectronic Quantum Architecture and Measurement (MOE), School of Physics, Beijing Institute of Technology, Beijing, 100081, China}
\affiliation{Beijing Key Lab of Nanophotonics \& Ultrafine Optoelectronic Systems, School of Physics, Beijing Institute of Technology, Beijing, 100081, China}

\author{Yugui Yao}
\email{ygyao@bit.edu.cn}
\affiliation{Centre for Quantum Physics, Key Laboratory of Advanced Optoelectronic Quantum Architecture and Measurement (MOE), School of Physics, Beijing Institute of Technology, Beijing, 100081, China}
\affiliation{Beijing Key Lab of Nanophotonics \& Ultrafine Optoelectronic Systems, School of Physics, Beijing Institute of Technology, Beijing, 100081, China}

\begin{abstract}
Using symmetry analysis and semiclassical Boltzmann equation, we theoretically explore the planar Hall effect (PHE) in three-dimensional materials.
We demonstrate that PHE is a general phenomenon that can occur in various systems regardless of band topology.
Both the Lorentz force and Berry curvature effects  can induce significant PHE, and the leading contributions of both effects   linearly depend on the electric and magnetic fields.
The Lorentz force and  Berry curvature PHE coefficient possess only antisymmetric and symmetric parts, respectively.
Both contributions respect  the same crystalline symmetry constraints but differ under time-reversal symmetry.
Remarkably, for topological Weyl semimetal, the Berry curvature PHE coefficient is a constant that does not depends on the Fermi energy, while the Lorentz force contribution linearly increases with the Fermi energy, resulting from the linear dispersion of the Weyl point.
Furthermore, we find that the PHE in topological nodal line semimetals is mainly induced by the Lorentz force, as the Berry curvature in these systems vanishes near the nodal line.
Our study not only highlights the significance of the Lorentz force in PHE, but also reveals its unique characteristics, which will be beneficial for determining the Lorentz force contribution experimentally.
\end{abstract}

\maketitle

\section{Introduction}
The study of magnetotransport has always attracted extensive attention in condensed matter physics.
Due to the Lorentz force, electrons experience  transverse force perpendicular to their moving direction under a  magnetic field.
Hence, for the systems with crossed electric and magnetic fields, it is easy to expect that there will exist a  transverse current perpendicular to both electric and magnetic fields, known as the ordinary  Hall effect \cite{ashcroftsolid, RevModPhys.82.1959}.
Besides, there are several other related phenomena, such as the anomalous Hall effect, spin Hall effect, and quantum (spin) Hall effect  \cite{RevModPhys.82.1539AHE, RevModPhys.87.1213spinHE, QHE_Klitzing, RevModPhys.67.357QHE, PhysRevLett.96.106802QSHE}.
All of these Hall effects are crucial  for both theoretical investigations and practical  applications \cite{RevModPhys.82.1539AHE, RevModPhys.87.1213spinHE, QHE_Klitzing, RevModPhys.67.357QHE, PhysRevLett.96.106802QSHE, 56910,1634415,DAUGHTON1999334}.

Interestingly, it has been discovered that a coplanar magnetic field and electric field can also induce a significant transverse current \cite{doi:10.1063/1.364592, doi:10.1063/1.123249, PhysRevLett.90.107201,PhysRevB.71.172401}.
Since the exerted electric field, magnetic field, and Hall current all lie in the same plane, this phenomenon is called as planar Hall effect (PHE) \cite{doi:10.1063/1.364592, doi:10.1063/1.123249, PhysRevLett.107.086603, PhysRevLett.90.107201, PhysRevB.71.172401, PhysRevB.101.041408, PhysRevB.102.241105,PhysRevLett.123.016801, PhysRevLett.126.256601, xiang2023intrinsic, wang2022theory, wang2023fieldinduced,PhysRevB.107.075131, PhysRevLett.130.126303, cao2022inplane}.
Compared to the ordinary Hall effect, it generally  is difficult  to develop an intuitive picture to understand the PHE.
Pal \textit{et. al.} \cite{PhysRevB.81.214438} investigated the longitudinal magnetoresistance induced by the Lorentz force and showed that  certain kinds of anisotropic spectrum may be  relevant to the PHE.

In the past decade, with the discovery of graphene and topological Weyl semimetal, the field of topological materials has undergone rapid development \cite{RevModPhys.83.1057, RevModPhys.90.015001, RevModPhys.93.025002}.
Besides the Weyl and Dirac points, the conduction and valence bands of crystals can form many different kinds of degeneracies around the Fermi energy, such as triple point, nodal line, and nodal surface \cite{RN1178, PhysRevX.6.031003, PhysRevB.93.241202, PhysRevB.97.115125,RN1450,YUScienceBulletin,PhysRevB.106.214309, runwu, PhysRevB.106.195129}.
Around these degeneracies, the electronic band generally exhibits significant  Berry curvature and non-trivial band topology \cite{RevModPhys.90.015001, RevModPhys.93.025002}, leading to various exotic properties \cite{PhysRevLett.117.077202, PhysRevLett.125.076801, PhysRevLett.119.206401, PhysRevX.10.041041, PhysRevB.97.125143,PhysRevB.103.205104,PhysRevLett.119.166601,PhysRevB.107.085146,PhysRevB.106.L081121,zhu2022thirdorder,PhysRevB.105.045118,PhysRevLett.127.277202}.
The Berry curvature also has an important influence on the transport properties of systems \cite{RevModPhys.82.1959}.
Nandy \textit{et. al.} \cite{PhysRevLett.119.176804}  have shown that the PHE can naturally appear in Weyl semimetals as a result of chiral anomaly, with the PHE conductivity depending quadratically on the magnetic field.
This quadratic behavior has also been reported in many other works while with different origins \cite{PhysRevB.100.235105,PhysRevB.105.205207}.
However,  in the regime of strong $B$ field, the chiral anomaly can lead to a linear dependence of the PHE on the  field \cite{PhysRevB.99.165146,PhysRevB.106.075139}.
Besides, for the systems  with lower symmetry, such as anisotropic  Weyl cone, the PHE conductivity also scales with the first order of  magnetic field  \cite{PhysRevB.99.115121,PhysRevB.98.205139,PhysRevB.105.205126}.The PHE  has been experimentally observed in many topological semimetals, such as $\text{PbTe}_2$, $\text{ZrTe}_5$, $\text{VAl}_3$ and $\text{WTe}_2$ \cite{PHE1,PHE2,PHE3,PHE4,PHE5,PHE6,PHE7,PHE8,PHE9,PHE10,sonika2023planar, li2023higher, Zhong_2023}.
However, there remains a lack of systematic investigation to show under which condition the PHE can  or cannot occur.
Besides, current research on PHE is generally based on Weyl semimetals and certain topological insulators \cite{PhysRevLett.119.176804,PhysRevB.98.205139,PhysRevB.102.205107, PhysRevB.105.205126,PhysRevB.105.205207,YADAV2022115444, PhysRevB.99.115121,SciPostPhysCore, PhysRevB.107.L081110,nandy2018berry,taskin2017planar,Wang_2021}.
In contrast, the nodal line naturally exhibits strong anisotropic spectrum \cite{YUScienceBulletin}, suggesting that for PHE,  the Lorentz force may play a more important role in nodal line systems than that in Weyl semimetals.

{
\global\long\def\arraystretch{1.4}%
\begin{table*}[htbp]
\caption{\label{tab} The constraints on $\boldsymbol{\chi}_{yx}^{\text{even}}$ and $\boldsymbol{\chi}_{yx}^{\text{odd}}$ for some representative symmetry operations. \textquotedblleft$\checkmark$\textquotedblright (\textquotedblleft$\bm{\times}$\textquotedblright ) means the element is symmetry allowed (forbidden) by the operation. }
\begin{ruledtabular}
\begin{tabular}{cccccccccccccccc}
& $\mathcal{T}$ & $\mathcal{P}$ & $\mathcal{PT}$ & $\mathcal{C}_{2x}$ & $\mathcal{C}_{2y}$ & $\mathcal{C}_{2z}$ & $\mathcal{M}_{x}$ & $\mathcal{M}_{y}$ & $\mathcal{M}_{z}$ & $\mathcal{C}_{3z}$ & $\mathcal{C}_{4z}$ & $\mathcal{C}_{6z}$ & $\mathcal{S}_{3z}$ & $\mathcal{S}_{4z}$ & $\mathcal{S}_{6z}$\tabularnewline
\hline
$\chi_{yx,x}^{\text{odd}}$ & $\bm\times$ & $\checkmark$ & $\bm\times$ & $\bm\times$ & $\checkmark$ & $\bm\times$ & $\bm\times$ & $\checkmark$ & $\bm\times$ & $\checkmark$ & $\bm\times$ & $\bm\times$ & $\bm\times$ & $\bm\times$ & $\checkmark$\tabularnewline
$\chi_{yx,x}^{\text{even}}$ & $\checkmark$ & $\checkmark$ & $\checkmark$ & $\bm\times$ & $\checkmark$ & $\bm\times$ & $\bm\times$ & $\checkmark$ & $\bm\times$ & $\checkmark$ & $\bm\times$ & $\bm\times$ & $\bm\times$ & $\bm\times$ & $\checkmark$\tabularnewline
\hline
$\chi_{yx,y}^{\text{odd}}$ & $\bm\times$ & $\checkmark$ & $\bm\times$ & $\checkmark$ & $\bm\times$ & $\bm\times$ & $\checkmark$ & $\bm\times$ & $\bm\times$ & $\checkmark$ & $\bm\times$ & $\bm\times$ & $\bm\times$ & $\bm\times$ & $\checkmark$\tabularnewline
$\chi_{yx,y}^{\text{even}}$ & $\checkmark$ & $\checkmark$ & $\checkmark$ & $\checkmark$ & $\bm\times$ & $\bm\times$ & $\checkmark$ & $\bm\times$ & $\bm\times$ & $\checkmark$ & $\bm\times$ & $\bm\times$ & $\bm\times$ & $\times$ & $\checkmark$\tabularnewline
\end{tabular}
\end{ruledtabular}
\end{table*}
}

In this work, we perform a systematic symmetry analysis of the  PHE conductivity, and show that linear PHE can exist in various systems with and without band topology.
Then via the semiclassical Boltzmann equation, we expand the PHE conductivity to the linear order of the magnetic field, and find that both the Lorentz force and Berry curvature can cause linear PHE.
Particularly,  the  Berry curvature contribution to linear PHE vanishes when the systems have time-reversal symmetry (${\cal{T}}$).
This means that even in topological semimetals, the  Lorentz force may dominate linear PHE, as long as the system has ${\cal{T}}$ symmetry.
For general cases,  both Berry curvature and Lorentz force contribute to PHE.
We find that for the Weyl point, the Berry curvature  (Lorentz force) contribution dominates linear PHE when the Fermi energy is close to (far  from) the Weyl point, as the former is a constant  that does not depends on the Fermi energy, while the latter linearly increases with the Fermi energy.
However, for the nodal line systems, the Lorentz force contribution  always dominates the linear PHE regardless of the position of  Fermi level.
Our results provide important insights into the PHE in topological semimetals, and are ready for experimental examination.

The paper is organized as follows. In Sec. \Rom{2}, we provide a systematic symmetry analysis of the PHE conductivity and summarize all the results in a table. In Sec. \Rom{3}, we derive the expressions for PHE conductivity based on the semiclassical Boltzmann transport theory. The specific numerical calculations of different models are described in Sec. \Rom{4}. Finally, Sec. \Rom{5} contains a brief summary and discussion.

\section{General analysis}
Without loss of generality, we assume that the exerted electric field $E$, magnetic field $B$ and Hall current all lie in the $x$-$y$ plane, and  the $x$ axis aligns with the electric field direction.
According to the linear response theory, the PHE conductivity  $\sigma_{yx}$  is obtained by \cite{RevModPhys.82.1539}
\begin{equation}
	j_{y} = \sigma_{yx}(B_x, B_y) E_x,
\end{equation}
where  $j_{y}$ is the  Hall current density, and $B_x=B \cos\theta$ and $B_y=B \sin \theta$ with $\theta$ denoting the  angle between $x$ axis and $B$ field.
When the magnetic field  is weak,  we can expand $\sigma_{yx}(B_x, B_y)$  in the powers of $B_{x(y)}$.
Up to the linear order, $\sigma_{yx}(B_x, B_y)$ can be approximately written as
\begin{equation}\label{sigma}
	\sigma_{yx} = \sigma_{yx}^{0}+\boldsymbol{\chi}_{yx}\cdot \boldsymbol{B}\equiv\sigma_{yx}^{0}+\chi_{yx,x} B_x+\chi_{yx,y} B_y.
\end{equation}
$\sigma_{yx}^{0}$ has no dependence on $B$, resulting  from the classical Drude conductivity and intrinsic anomalous Hall conductivity, while $\boldsymbol{\chi}_{yx}\cdot \boldsymbol{B}$ corresponds to the magnetoconductivity induced by the magnetic field.
Generally, the  coefficient $\boldsymbol{\chi}_{yx}$ in Eq. (\ref{sigma}) is a function of the  relaxation time $\tau$.
Hence, we can further  divide it into two parts,
\begin{equation}
	\boldsymbol{\chi}_{yx} = \boldsymbol{\chi}_{yx}^{\text{even}}+\boldsymbol{\chi}_{yx}^{\text{odd}},
\end{equation}
where $\boldsymbol{\chi}_{yx}^{\text{even}}$ ($\boldsymbol{\chi}_{yx}^{\text{odd}}$) is an even (odd) function of $\tau$.
Then, the  PHE conductivity can be rewritten as
\begin{equation}
	\sigma_{yx} = \sigma_{yx}^{0}+\sigma_{yx}^{\text{even}}+\sigma_{yx}^{\text{odd}},
\end{equation}
with $\sigma_{yx}^{\text{even}}=\boldsymbol{\chi}_{yx}^{\text{even}}\cdot \boldsymbol{B}$ and  $\sigma_{yx}^{\text{odd}}=\boldsymbol{\chi}_{yx}^{\text{odd}}\cdot \boldsymbol{B}$.

The expressions of  $\boldsymbol{\chi}_{yx}^{\text{even}}$ and $\boldsymbol{\chi}_{yx}^{\text{odd}}$ should respect  the magnetic point group symmetry of  systems. It should be stressed  that  the relaxation  time $\tau$ in  $\boldsymbol{\chi}_{yx}$  reverses its sign under $\cal{T}$ symmetry and the magnetic symmetry operators  $\cal{OT}$ with $\cal{O}$ a spatial operator \cite{PhysRevX.10.041041,RevModPhys.82.1539}.
We summarize the behaviors of $\boldsymbol{\chi}_{yx}^{\text{even(odd)}}$ under different symmetry operations in Table \ref{tab}.
One finds that $\boldsymbol{\chi}_{yx}^{\text{even}}$ and $\boldsymbol{\chi}_{yx}^{\text{odd}}$ respect the same spatial symmetry constraints, and only a few spatial operations like $\mathcal C_{2z}$ and $\mathcal M_{z}$ can completely suppress them (Notice that ${\cal{C}}_{4z}^2={\cal{C}}_{6z}^3={\cal{S}}_{4z}^2={\cal{C}}_{2z}$ and ${\cal{S}}_{3z}^3=({\cal{C}}_{6z}{\cal{P}})^3={\cal{M}}_{z}$.).
In addition, $\boldsymbol{\chi}_{yx}^{\text{odd}}$ vanishes in the nonmagnetic systems.
Because the symmetry analysis is irrelevant to  the band topology, the results in Table \ref{tab} clearly show that the PHE can be realized in  a wide variety of systems regardless of the band topology.

\section{Semiclassical Theory of  PHE conductivity}
We then study the microscopic origin of the PHE conductivity $\sigma_{yx}$  based on the semiclassical Boltzmann transport theory.
In the presence of Berry curvature, the Bloch electrons under weak electric and magnetic fields can be described by the following semiclassical equations of motion \cite{RevModPhys.82.1959,PhysRevLett.95.137204},
\begin{align}
	\dot {\bm r} &=  D(\bm B,\bm{\Omega_k}) \left[\bm{\tilde v} + \frac e \hbar \bm E \times \bm{\Omega_k} + \frac e \hbar (\bm{\tilde v} \cdot \bm{\Omega_k})\bm B\right], \\
	\dot {\bm k} &=   D(\bm B,\bm{\Omega_k})\left[ -\frac e\hbar \bm E -\frac e\hbar \bm{\tilde v} \times \bm B -\frac {e^2}{\hbar^2} (\bm E \cdot \bm B)\bm{\Omega_k}  \right],
\end{align}
where $\dot {\bm r}$ and $\dot {\bm k}$ are the time derivatives of position $\bm r$ and wave vector $\bm k$,  $D(\bm B,\bm{\Omega_k}) \equiv [1 + e(\bm B \cdot \bm{\Omega_k})/\hbar]^{-1}$ is the modification of phase space volume, $\bm{\Omega_k}=-2\epsilon_{\alpha\beta\gamma}\text{Im}\langle \partial_{k_\alpha} u(\bm k) |\partial_{k_\beta} u(\bm k) \rangle$  denotes the Berry curvature with $\epsilon_{\alpha\beta\gamma}$  the Levi-Civita tensor, $u(\bm k)$  the cell-periodic part of Bloch function, and  $\bm{\tilde v} = \partial_{\bm k} \tilde\varepsilon/\hbar $ is the  velocity of electrons.
$\tilde\varepsilon(\bm{k}) = \varepsilon(\bm k) - \bm m \cdot \bm B$, $\varepsilon(\bm k)$ is the band dispersion and $\bm m$ is the orbital magnetic moment results from the semiclassical self-rotation of the electron wave packet \cite{RevModPhys.82.1959}.
 The electron charge is taken as $-e$ (i.e., $e > 0$).

The current density is given by \cite{ashcroftsolid,RevModPhys.82.1959}
\begin{equation}\label{curr}
	\bm J = -e \int \frac{\mathrm d^3k}{(2\pi)^3}D^{-1}\dot{\bm r} f_{\bm k}(\bm r,t),
\end{equation}
with  $f_{\bm k}(\bm r,t)$  the  distribution function, which can be determined by solving the semiclassical Boltzmann transport equation,
\begin{equation} \label{Boltz}
	\frac{\partial f_k}{\partial t} + \dot {\bm r}\cdot\frac{\partial f_k}{\partial \bm r} + \dot {\bm k}\cdot\frac{\partial f_k}{\partial \bm k} = \left(\frac{\partial f_k}{\partial t} \right)_\text{coll}.	
\end{equation}
The right side of Eq. (\ref{Boltz}) is known as the collision integral and generally can be treated by relaxation time approximation \cite{ashcroftsolid}.
Then, for a homogeneous system in the steady state, where $\partial f / \partial t$ and $\partial f / \partial\bm r$ vanish, Eq. (\ref{Boltz}) is simplified as
\begin{equation}\label{Boltz2}
	\dot {\bm k}\cdot\frac{\partial f_k}{\partial \bm k} = -\frac{f_k-f_k^{0}}{\tau}\equiv -\frac{g_k}{\tau},
\end{equation}
with  $f_k^{0}$  the original Fermi-Dirac distribution function and $\tau$ the relaxation time.
In the following calculations, we take the relaxation time  $\tau$ as 0.1 ps \cite{PhysRevLett.119.176804,PhysRevB.97.161404,shekhar2015extremely,PhysRevMaterials.4.034201}.
Notice that $g_k$ vanishes  when $E=0$. Hence, for a weak $E$ field, we can approximately  assume $g_k={\bm \Gamma}\cdot{ \bm E}$,
where ${\bm \Gamma}$ is a  function to be determined.
Substituting the expression of  $g_k$  into Eq. (\ref{Boltz2}), and keeping only the linear order of  $\bm E$ and $\bm B$ (as well as their combination $E_{i}B_{j}$), ${\bm \Gamma}$ and  the distribution function $f_k$  can be determined.
Then, the expression of current density is obtained as
\begin{eqnarray}
&&\bm{J} =  e^{2}\int [d \bm{k} ]\left[\tau\hbar \boldsymbol{v}(\boldsymbol{v}\cdot\boldsymbol{E})\partial_{\varepsilon}f_k^0-\left(\boldsymbol{E}\times\boldsymbol{\Omega}_{\boldsymbol{k}}\right)f_k^0\right] \nonumber\\
 &  &-e^2\tau \int [d \bm{k} ] \left[ \bm v \left(\partial_{\bm k}(\bm m \cdot {\bm B}) \cdot \bm E \right) + (\bm v \cdot \bm E)\partial_{\bm k}(\bm m \cdot {\bm B}) \right]\partial_{\varepsilon}f_k^0 \nonumber\\
 &  &-e^{3}\tau \int [d \bm{k} ]\boldsymbol{v}(\boldsymbol{v}\cdot\boldsymbol{E})(\bm B \cdot \bm{\Omega_k})\partial_{\varepsilon}f_k^0 \nonumber\\
 &  & +e^{3}\tau\int [d \bm{k} ]\left[\boldsymbol{B}(\boldsymbol{v}\cdot\boldsymbol{E})+\boldsymbol{v}(\boldsymbol{B}\cdot\boldsymbol{E})\right](\boldsymbol{v}\cdot\boldsymbol{\Omega}_{\boldsymbol{k}})\partial_{\varepsilon}f_k^0\nonumber\\
 &  & +e^{3}\tau^{2}\int [d \bm{k} ] \boldsymbol{v}\left[(\boldsymbol{v}\times\boldsymbol{B})\cdot\partial_{\boldsymbol{k}}(\boldsymbol{v}\cdot\boldsymbol{E})\right]\partial_{\varepsilon}f_k^0,
\end{eqnarray}
with $\int [d \bm{k} ]\equiv -\frac{1}{(2\pi)^3\hbar} \int \mathrm{d^3}k $ and $\bm{v} = \partial_{\bm k} \varepsilon/\hbar $.
This result is consistent with the previous works \cite{PhysRevB.81.214438,PhysRevB.99.115121} but it captures all the terms up to order $O(E)$ and $O(EB)$.
Consequently, the expressions of the  PHE conductivity $\sigma_{yx} = \sigma_{yx}^{0}+\sigma_{yx}^{\text{even}}+\sigma_{yx}^{\text{odd}}$ are
\begin{eqnarray}
	\sigma^{0}_{yx}&=& e^2\int [d \bm{k} ] (\tau\hbar  v_x v_y \frac{\partial f_k^0}{\partial \varepsilon} -\Omega_{k}^zf_k^0),
\end{eqnarray}
which are the classical Drude conductivity and intrinsic anomalous Hall conductivity,
and
\begin{eqnarray}
\sigma^{\text{even}}_{yx}&=&  e^3\tau^2 \int [d \bm{k} ]\frac{\partial f_k^0}{\partial \varepsilon} v_y(\bm v\times\bm B) \cdot \frac{\partial v_x}{\partial \bm k}, \label{seven} \\
\sigma^{\text{odd}}_{yx}&=&   e^3\tau \int [d \bm{k} ]  \frac{\partial f_k^0}{\partial \varepsilon} \Bigl[ -(\bm B \cdot \bm{\Omega}_{\bm{k}})v_xv_y \nonumber \\
 & &    -\frac{v_y}{e}\partial_{k_x}(\bm m \cdot \bm B) - \frac{v_x}{e}\partial_{k_y}(\bm m \cdot {\bm B}) \nonumber\\
 & &    + (B_x v_y+ B_y v_x) (\bm v \cdot \bm{\Omega}_{\bm{k}}) \Bigr].  \label{sodd}
\end{eqnarray}

Clearly,  $\sigma_{yx}^{\text{odd}}$  is induced by the Berry curvature and is an odd function of $\tau$, as it  is in direct proportion to $\tau$.
In contrast,  $\sigma_{yx}^{\text{even}}$ describes the   Lorentz force contribution to PHE conductivity and is an even function of $\tau$.
This means that under ${\cal{T}}$ symmetry, $\sigma_{yx}^{\text{even}}$ is invariant while $\sigma_{yx}^{\text{odd}}$ should reverse its sign. Hence, for the systems with ${\cal{T}}$ symmetry, $\sigma_{yx}^{\text{odd}}$ vanishes and the leading order of the Berry curvature contribution to PHE is quadratic.
Remarkably, we find that $\sigma_{yx}^{\text{odd}}$ is symmetric in its two indices, namely $\sigma_{yx}^{\text{odd}}= \sigma_{xy}^{\text{odd}}$.
Moreover, both  $\sigma_{yx}^{\text{odd}}$ and $\sigma_{yx}^{\text{even}}$ have the term of $\partial f_k^0 /\partial \varepsilon$, indicating that only the electronic states around the Fermi level have contribution to them.

\section{Model Studies}
\subsection{Single-band model}
\begin{figure}
\begin{centering}
\includegraphics[width=\linewidth]{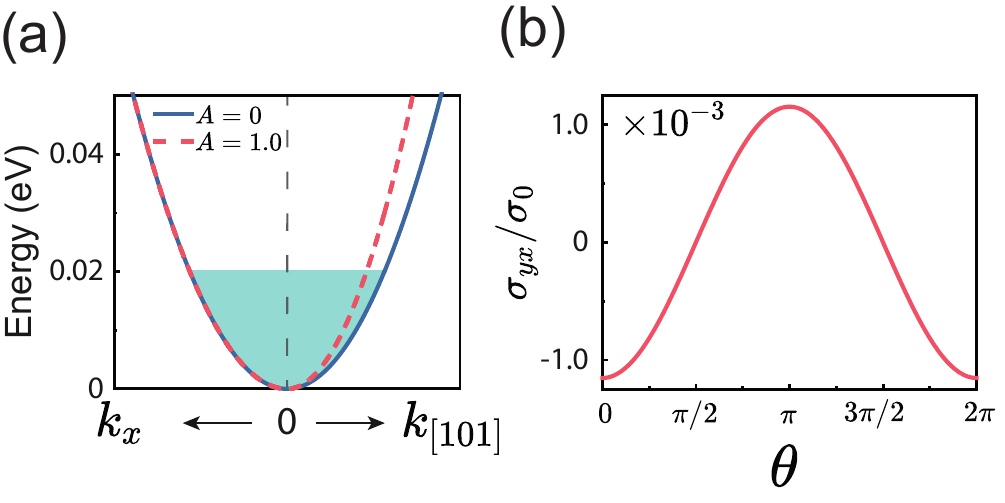}
\par\end{centering}
\caption{(a) Energy spectrum of the single-band model (\ref{model1}) without (with) anisotropic effect. (b) Calculated  PHC versus the angle $\theta$ for model (\ref{model1}), using Fermi level $E_\text{F} = 20 \ \text{meV}$ and the magnetic field $B = 0.5 \ \mathrm{T}$. In the calculations, we take the model parameter $m = 0.5 \ \mathrm{eV^{-1}\cdot\AA^{-2}}$, $A = 1.0 \ \mathrm{eV\cdot \AA^2}$ and the PHC is normalized by the corresponding longitudinal conductivity without magnetic field [i.e. $\sigma_{0} = \sigma_{xx}(B = 0; E_\text{F} = 20 \  \text{meV})$]. \label{Fig_1}}
\end{figure}

To unambiguously demonstrate  that the PHE can be induced by Lorentz force, we investigate a  single-band model, where the Berry curvature is definitely zero.
Consider a simple  model that meets the symmetry requirements listed in Table \ref{tab}, for which the  Hamiltonian may be written as
\begin{align} \label{model1}
	{\cal{H}}_{1} = \frac{k^2}{2m} + A k_x k_z,
\end{align}
where $k^2 = k_x^2 + k_y^2 + k_z^2$, $m$ is the  effective mass of electron and the parameter $A$ is a real number, denoting the anisotropy of the system [see Fig. \ref{Fig_1}(a)].
One can check that the model (\ref{model1}) breaks the $\mathcal C_{2z}$ symmetry when $A$ is finite.
Hence, PHE can (not) be realized in it when $A\neq0$ ($A=0$).
As shown in Fig. \ref{Fig_1}(b), this extremely simple model indeed produces a significant PHE signal with an angular dependence  proportional to $\cos\theta$.
Since the anisotropy is a common feature in  real materials, one can expect that PHE in most real materials can be significant and has a period of $2\pi$  by varying  the direction of magnetic field in the $x$-$y$ plane.

\begin{figure}
	\begin{centering}
	\includegraphics[width=\linewidth]{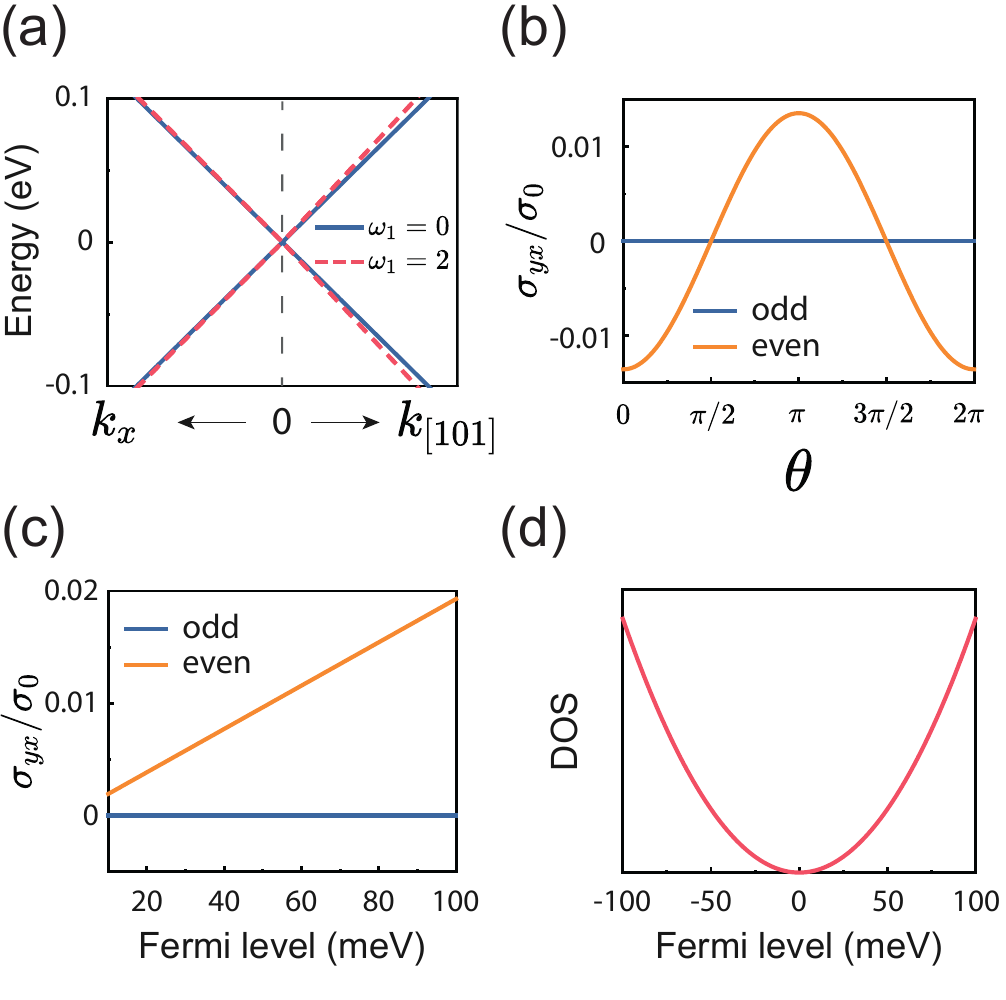}
	\par\end{centering}
	\caption{(a) Band structure of Weyl model (\ref{ham2}). (b) Calculated $\sigma^{\text{odd}}_{yx}$ and $\sigma^{\text{even}}_{yx}$ as  functions of angle $\theta$, using $E_\text{F} = 50 \ \text{meV}$ and $B = 0.05 \ \mathrm{T}$. (c) Variation of $\sigma^{\text{odd}}_{yx}$ and $\sigma^{\text{even}}_{yx}$ with Fermi level with $\theta = 3\pi/4$. (d) The density of states (DOS) of Weyl model (\ref{ham2}). In the calculations, we take $\omega_1 = 2.0 \ \text{eV}\cdot\text\AA$, $v_1 = 12.0 \ \text{eV}\cdot\text\AA$ and the  conductivities in (b) and (c) are normalized by  $\sigma_{0} = \sigma_{xx}(B = 0; E_\text{F} = 100 \ \text{meV})$. \label{Fig_2}}
\end{figure}

\subsection{Weyl semimetals}
Although the PHE in Weyl semimetals has been studied in previous works \cite{PhysRevLett.119.176804, PhysRevB.105.205207, PhysRevB.99.115121,PhysRevB.98.205139,PhysRevB.105.205126}, here we focus on the  competition between the Berry curvature and the Lorentz force
contributions to PHE, and show under what conditions the Berry curvature or  the Lorentz force contribution will dominate PHE.
For ideal Weyl point described by $\bm  k \cdot \bm\sigma$ ($\bm\sigma$s are  Pauli matrices), both $\sigma_{yx}^{\text{odd}}$ and $\sigma_{yx}^{\text{even}}$ are zero, due to the high symmetry of this model.
Consequently, the chiral anomaly contribution dominates PHE in conventional topological Weyl semimetals, resulting in PHE having a quadratic dependence on the  magnetic field and an angular dependence with a period of $\pi$ \cite{PhysRevLett.119.176804}.

Breaking $\mathcal C_{2z}$ but keeping ${\cal{T}}$ symmetry leads to the vanishing of $\sigma_{yx}^{\text{odd}}$, while the Lorentz force contribution $\sigma_{yx}^{\text{even}}$ will appear and dominate the PHE in the limit of weak $B$-field.
To demonstrate this directly,  we consider the following Weyl Hamiltonian
\begin{align}\label{ham2}
{\cal{H}}_{2} = w_1 k_x\sigma_z+v_1 \bm k \cdot \bm\sigma,
\end{align}
where  $w_1$ and $v_1$ are  real  parameters.
The energy dispersion of this two-band model is $\varepsilon_{\pm}(\bm k) = \pm \sqrt{k^2v_1^2 + k_x^2w_1^2 + 2k_xk_z v_1 w_1}$, where $\pm$ denotes the conduction and valence bands, respectively. From the band structure shown in Fig. \ref{Fig_2}(a), one observes that the additional term $w_1 k_x\sigma_z$ increases the slope along [101] direction and changes the shape of the Fermi surface from a sphere to an ellipsoid.

The calculated results of PHE based on Hamiltonian ${\cal{H}}_{2}$ (\ref{ham2}) are shown  in  Fig. \ref{Fig_2}.
One observes that  $\sigma_{yx}^{\text{odd}}$ is always zero regardless of the direction of $B$-field as guaranteed by the ${\cal{T}}$ symmetry, while  $\sigma_{yx}^{\text{even}}$ is finite except  ${\bm B} \perp {\bm E}$ and becomes largest when  ${\bm B} || {\bm E}$. The period of  $\sigma_{yx}^{\text{even}}$ in $\theta$ is $2\pi$.
Since the density of state of the Weyl system (\ref{ham2}) increases with  the Fermi energy [Fig. \ref{Fig_2}(d)], one can expect that $\sigma_{yx}^{\text{even}}$ would become more and more significant by raising  the Fermi energy.
The calculated $\sigma_{yx}^{\text{even}}$ as a function of $E_F$ is shown in Fig. \ref{Fig_2}(c), where a linear increase of  $\sigma_{yx}^{\text{even}}$  is observed.
This linear dependence of $\sigma_{yx}^{\text{even}}$ on $E_\text{F}$  is guaranteed by the linear Hamiltonian of the Weyl system, and can be easily obtained by the  scaling analysis proposed by Cao et. al. \cite{cao2022lowfrequency}.
Specifically, let us consider a scaling transformation in momentum and energy: $\bm k \rightarrow \lambda \bm k$ and $E_F \rightarrow \lambda E_F $, where  $\lambda$ is a real number. Due to the linear Hamiltonian, one has ${\cal{H}}_2(\lambda \bm k)=\lambda{\cal{H}}_2( \bm k)$ and $\varepsilon(\lambda \bm k)=\lambda \varepsilon(\bm k)$.
However, the velocity and  the eigenstates of system are invarient under the scaling, i.e. $v(\lambda \bm k)=v(\bm k)$ and $|u_{n,\lambda \bm k}\rangle=|u_{n,\bm k}\rangle$. Thus, for Eq. (\ref{seven}), we find that  $\sigma^{\text{even}}_{yx}(\lambda E_F)=\lambda \sigma^{\text{even}}_{yx}(E_F)$, indicating $\sigma^{\text{even}}_{yx}(E_F)\propto E_F$.

\begin{figure}
	\begin{centering}
		\includegraphics[width=\linewidth]{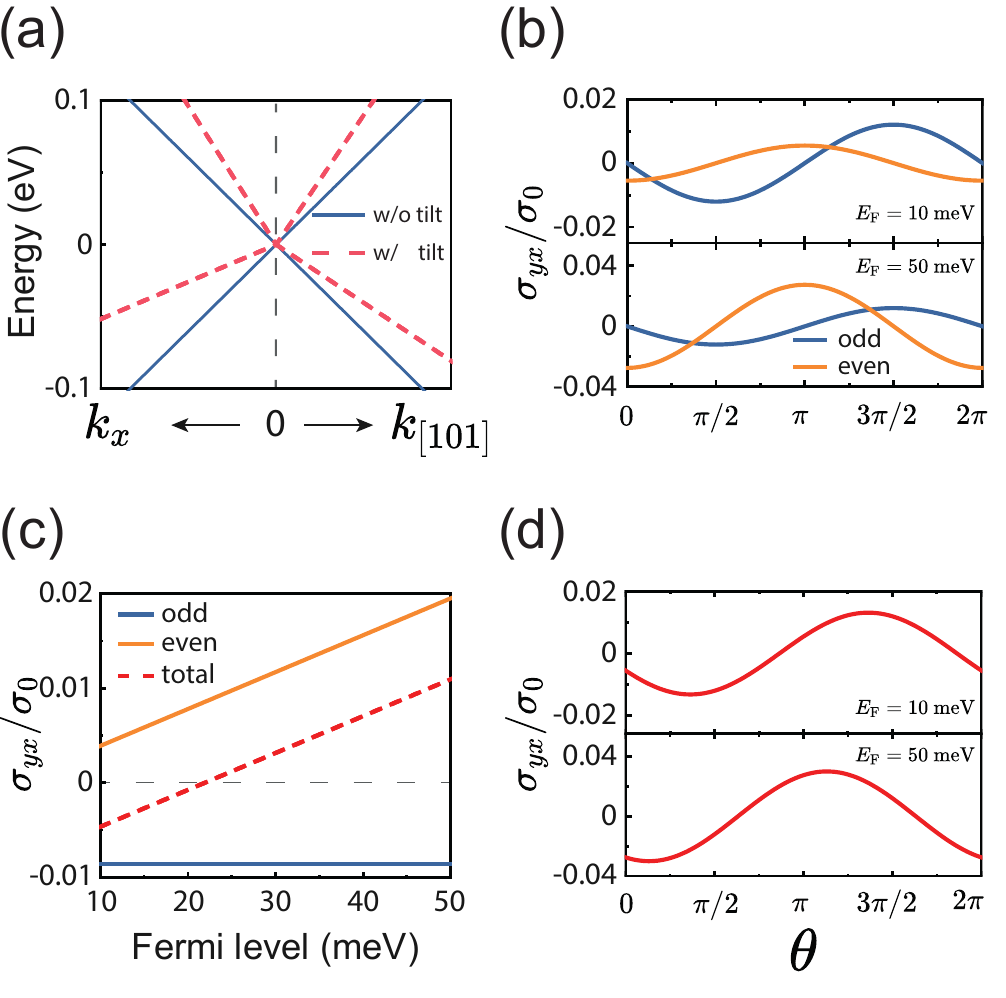}
		\par\end{centering}
	\caption{(a) Band structure of Weyl model (\ref{ham3}) without (with) tilt term. (b) The competition between $\sigma^{\text{odd}}_{yx}$ and $\sigma^{\text{even}}_{yx}$ under different Fermi level. (c) Variation of $\sigma^{\text{odd}}_{yx}$, $\sigma^{\text{even}}_{yx}$ and the total PHC $\sigma^{\text{total}}_{yx}$ with Fermi level for $\theta = 3\pi/4$. (d) The calculated total PHC versus angle $\theta$ with $E_\text{F} = 10 \ \text{and} \ 50 \ \text{meV}$. In the calculations, we take magnetic filed $B = 0.05 \ \text{T}$, model parameter $\omega_1 = 1.0 \ \text{eV}\cdot\text\AA$, $\omega_2 = 7.0 \ \text{eV}\cdot\text\AA$ , $v_1 = 12.0 \ \text{eV}\cdot\text\AA$ and the conductivities in (b), (c) and (d) are normalized by  $\sigma_{0} = \sigma_{xx}(B = 0; E_\text{F} = 50 \ \text{meV})$.\label{Fig_3}}
\end{figure}

When both $\mathcal C_{2z}$ and ${\cal{T}}$ symmetries of systems are broken,  $\sigma_{yx}^{\text{odd}}$ and $\sigma_{yx}^{\text{even}}$  become finite.
However, because the Berry curvature  is significantly reduced by increasing Fermi energy in Weyl semimetals, it can be expected that $\sigma_{yx}^{\text{odd}}$ would not linearly increase with  the Fermi energy.
In fact, according to the aforementioned  scaling analysis, one can find that
$\sigma^{\text{odd}}_{yx}(\lambda E_F)= \sigma^{\text{odd}}_{yx}(E_F)$ due to ${\bm{\Omega}_{\lambda \bm{k}}}=\lambda^{-2} {\bm{\Omega}_{ \bm{k}}}$ and ${\bm m}_{\lambda \bm{k}} = \lambda^{-1} {\bm m}_{\bm{k}}$.
Hence, the $\sigma_{yx}^{\text{odd}}$ is a constant that does not depend on $E_\text{F}$.
Thus, the competition between $\sigma_{yx}^{\text{odd}}$ and $\sigma_{yx}^{\text{even}}$  may lead a sign reversal of the total  PHC conductivity by increasing the $E_F$.

The Weyl Hamiltonian breaking $\mathcal C_{2z}$ and ${\cal{T}}$ may be written as
\begin{align}\label{ham3}
{\cal{H}}_{3} = w_2 k_x  +w_1 k_x\sigma_z+v_1 \bm k \cdot \bm\sigma,
\end{align}
where  the additional term $w_2 k_x$ breaks ${\cal{T}}$ symmetry.
For this Weyl model, we numerically evaluate $\sigma_{yx}^{\text{odd}}$ and $\sigma_{yx}^{\text{even}}$ as  functions of $\theta$ and $E_\text{F}$, and the obtained results are plotted in Fig. \ref{Fig_3}.
Again,  $\sigma_{yx}^{\text{odd}}$ and $\sigma_{yx}^{\text{even}}$ vary with the direction of $B$-field in the period of $2\pi$  due to the linear dependence on ${\bm B}$.
Remarkably, we find that for electron doping  ($E_\text{F}>0$), the  $\sigma_{yx}^{\text{even}}$ is always positive and linearly increases with  $E_\text{F}$, while $\sigma_{yx}^{\text{odd}}$  is  a negative constant.
Thus, the total PHE conductivity is negative for small $E_\text{F}$ but changes its sign when  $E_\text{F}$ becomes large.
This sign change of the total PHC may be detected in experiments.

\subsection{Nodal line semimetals}
At last, we study the PHE in topological nodal line semimetals.
Numbers of real materials have been predicted to be topological nodal line semimetals, and most of them exhibit ${\cal{T}}$ symmetry \cite{PhysRevB.84.235126, PhysRevB.92.045108,PhysRevLett.115.036807,Fang_2016,PhysRevB.96.081106,PhysRevB.97.245148,runwu}.
With the above symmetry analysis, we know that in these nonmagnetic nodal line semimetals, the PHE is mainly induced by the Lorentz force when the fields are weak.
Moreover, for the nodal line protected by ${\cal{PT}}$ symmetry (${\cal{P}}$ is the spatial inversion symmetry), the Lorentz force contribution also dominates the PHE, as  ${\cal{PT}}$ symmetry  guarantees the Berry curvature of systems to vanish.

\begin{figure}
	\centering
	\includegraphics[width=\linewidth]{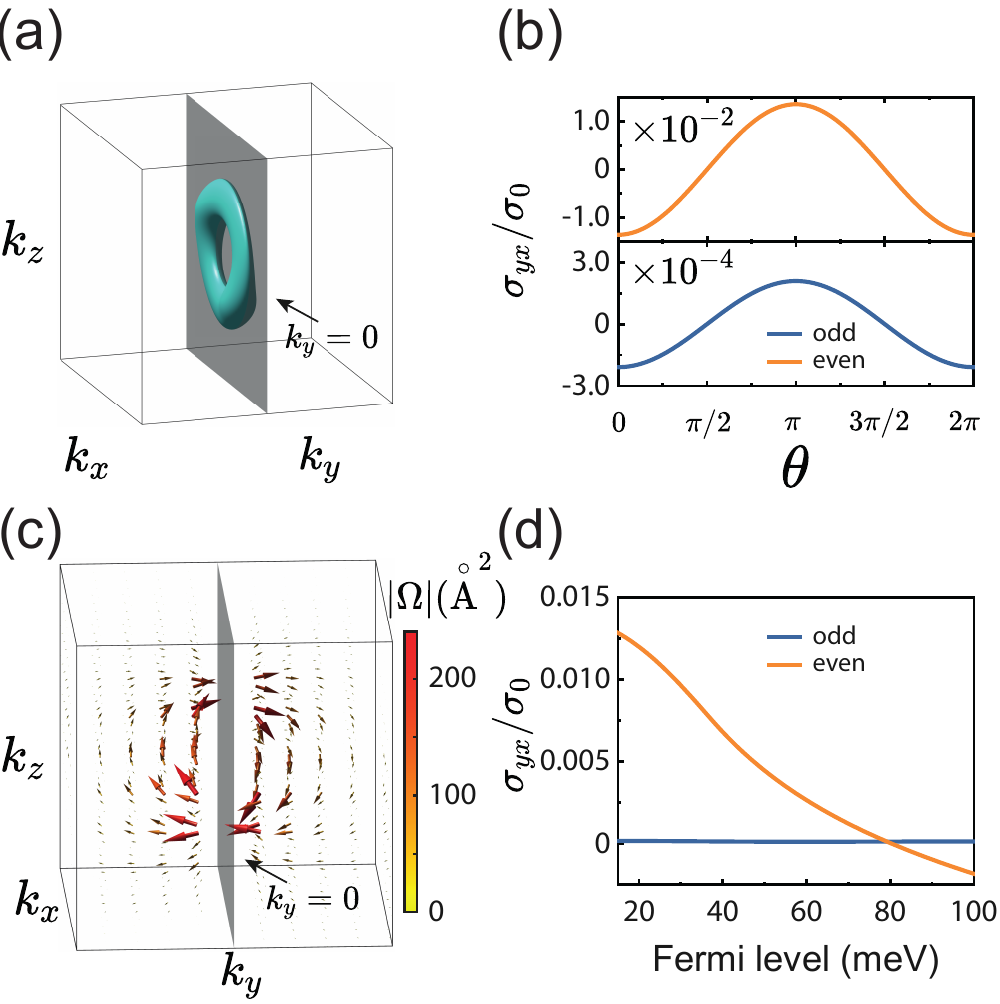}
	\caption{(a) Side view of model (\ref{hamNL})'s the Fermi surface. (b) The normalized $\sigma^{\text{odd}}_{yx}$ and $\sigma^\text{even}_{yx}$ as functions of angle $\theta$. (c) The Berry curvature distribution of the nodal line model (\ref{hamNL}) in the Brillouin zone (BZ). (d) Variation of $\sigma^{\text{odd}}_{yx}$ and $\sigma^\text{even}_{yx}$ with Fermi level for $\theta = 3\pi/4$. $\sigma^{\text{odd}}_{yx}$ is several  orders of magnitude smaller than $\sigma^{\text{even}}_{yx}$.
	In (a) and (b), we take Fermi level $E_\text{F} = 30 \  \text{meV}$. In all calculations, we take $B = 0.5 \ \text{T}$, $c = 1.0 \ \text{eV}\cdot\text{\AA}^2$, $m_0 = 0.04 \ \text{eV}$, $m_1 = 5.0 \ \text{eV}\cdot\text{\AA}^2$, $v_y = 0.9 \ \text{eV}\cdot\text{\AA}$, $\omega_1 = \omega_2 = 14.0 \ \text{eV}\cdot\text{\AA}^2$, $\omega_3 = 30.0 \ \text{eV}\cdot\text{\AA}^2$ and the conductivities in (b) and (d) are normalized by  $\sigma_{0} = \sigma_{xx}(B = 0; E_\text{F} = 30 \ \text{meV})$.\label{Fig_4}}
\end{figure}

Here, we  consider a nodal line model that breaks both ${\cal{T}}$ and ${\cal{PT}}$  symmetries to explore the competition between Berry curvature  and  Lorentz force effects.
For a concrete example, we take a  nodal line  with a mirror symmetry  $\mathcal M_y$, for which a general Hamiltonian  may be written as
\begin{align}\label{hamNL}
	{\cal{H}}_4 = &c k^2 + (m_0 - m_1 k^2)\sigma_z +v_yk_y\sigma_y \notag\\
	    & + (w_1k_x + w_2k_z + w_3k_x^2)k_y\sigma_x,
\end{align}
where  $c, m_0, m_1, v_y, \omega_1, \omega_2$ and  $\omega_3$ are real parameters. 
When $m_0m_1<0$, we have a mirror-protected nodal line lying in the $k_y = 0$ plane, as shown in Fig. \ref{Fig_4}(a).

In sharp contrast to the Weyl semimetals, we find that the Berry curvature  contribution to PHE ($\sigma_{yx}^{\text{odd}}$) in nodal line semimetals is several  orders of magnitude smaller than the Lorentz force contribution ($\sigma_{yx}^{\text{even}}$), as shown in Fig. \ref{Fig_4}(b) and \ref{Fig_4}(d). Further calculations reveal that the strong suppression of the Berry curvature contribution can be attributed to the special distribution of the Berry curvature of  the nodal line.

Generally, one can expect that the Berry curvature is significant around the band degeneracies, as it is  inversely proportional to the square of  band gap \cite{RevModPhys.82.1959}.
However, this is not the case for the mirror-protected nodal line. Since the model (\ref{hamNL}) has $\mathcal M_y$ symmetry and the nodal line  resides in $k_y=0$ plane, we study the  distribution of the Berry curvature of this system in  $k_y=0$ plane.
Because the Berry curvature ${\bm \Omega}({\bm k})$ is a pseudovector, only the $y$-component of ${\bm \Omega}({\bm k})$ can be nonzero in a  $\mathcal M_y$-invariant plane.
Hence, one has $\Omega_x=\Omega_z=0$ in the $k_y=0$ plane.
To calculate  $\Omega_y({k_y=0})$, we consider a 2D system, for which the  Hamiltonian is
\begin{align}\label{ham2DNL}
	&H_{2D}(k_x,k_z)=  {\cal{H}}_4(k_x,k_y=0,k_z) \notag\\
	    &= c k_x^2+c k_z^2 + (m_0 - m_1k_x^2-m_1k_z^2)\sigma_z.
\end{align}
A key feature  of this 2D Hamiltonian is that the two bands in it are decoupled, indicating that the Berry curvature  in this system must be zero, i.e. $\Omega_y({k_y=0})=0$.
Therefore, while the nodal line may exhibit sizable Berry curvature in momentum space, it features vanishing rather than divergent Berry curvature around the band degeneracy, i.e. the nodal line  [see Fig. \ref{Fig_4}(c)]. This  is completely different from the case in  Weyl point \cite{RevModPhys.90.015001}.

We also study the behaviors of $\sigma_{yx}^{\text{odd}}$ and  $\sigma_{yx}^{\text{even}}$ by varying $E_\text{F}$.
As shown in Fig. \ref{Fig_4}(d), although the Berry curvature contribution increases with Fermi energy, it is always much smaller than the Lorentz force contribution.
Another difference between Weyl points and nodal lines is that the Lorentz force contribution ($\sigma_{yx}^{\text{even}}$) in nodal line models may decrease when Fermi energy increases [see Fig. \ref{Fig_4}(d)], because the velocity of electrons around nodal line has a strong dependence on momentum and Fermi energy, and some of its components decrease when  the Fermi energy increases.

\begin{figure}
	\centering
	\includegraphics[width=\linewidth]{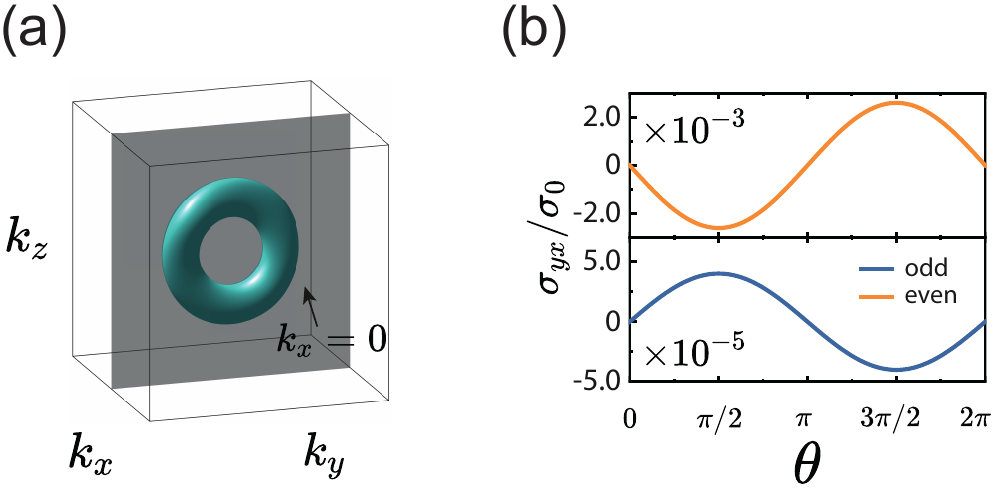}
	\caption{(a) The side view of Fermi surface of rotated nodal line model (\ref{hamNL}) with $E_\mathrm{F} = 30 \ \mathrm{meV}$. (b) The normalized $\sigma^{\text{odd}}_{yx}$ and $\sigma^\text{even}_{yx}$ as a function of angle $\theta$. The parameter is taken the same as Fig. (\ref{Fig_4})\label{Fig_5}}
\end{figure}

Additionally, the nodal line  is expected to exhibit strong direction-dependent transport behavior due to its highly anisotropic band dispersion.
We rotate the nodal line by $90 ^{\circ}$ around the $z$-axis and  the rotated Fermi surface is depicted in Fig. \ref{Fig_5}(a).
The  calculated PHE conductivity $\sigma_{yx}$ for this rotated nodal line is shown in Fig. \ref{Fig_5}(b)
As expected, the $\sigma_{yx}$ is still  dominated by the contribution from the Lorentz force.
The lower amplitude of  $\sigma^{\text{even}}_{yx}$ in Fig. \ref{Fig_5}(b) is ascribed to the higher longitudinal
conductivity $\sigma_{xx}(B = 0)$.

\section{DISCUSSION AND CONCLUSION}
In this work, we have studied the PHE in topological Weyl and nodal line semimetals. Both Lorentz force and Berry curvature contributions to PHE are discussed. We demonstrated that these two contributions respect the same crystalline symmetry constraints and are entirely suppressed by only a few symmetries.
However, their responses to ${\cal{T}}$ symmetry are opposite.
Our results indicate that the PHE can occur in a wide variety of systems with and without band topology.
Remarkably, we show the Lorentz force contribution can dominates the PHE in various topological systems, including Weyl semimetals with   ${\cal{T}}$ symmetry and topological nodal line semimetals.

In experimentation, the angular dependence and the order of relaxation time in  PHE  can be directly examined using magnetotransport measurements \cite{RN2811,RN2812}.
As discussed above, the Lorentz force PHE conductivity is directly proportional to $\bm{B}$ and $\tau^2$, while the Berry curvature PHE conductivity (including the PHE conductivity induced by chiral anomaly) is proportional to the odd order of  $\tau$.
Such unique behavior could be beneficial for experimentally identifying the  Lorentz force contribution to PHE.

Generally, both the Lorentz force and the Berry curvature can also induce finite PHE in topological nodal surface semimetals, which is protected by the presence of a two-fold screw rotation and ${\cal{T}}$ symmetry \cite{PhysRevB.97.115125}.
Compared with the Weyl points and the nodal lines, the anisotropy of the nodal surface is the strongest.
Particularly, the topological charge of the nodal surface can be any integer depending on the material details \cite{PhysRevB.100.041118, doi:10.1126/sciadv.aav2360}.
This means that the strength of the Berry curvature around the nodal surface can be tuned by model parameters.
Hence, the competition between the Berry curvature and the Lorentz force contributions to PHE in the nodal surface would be interesting, and varies with the model of the nodal surface.

\section{Acknowledgement}
The authors thank J. Xun for helpful discussions.
This work is supported the NSF of China (Grants Nos. 12004035, 12234003 and  12061131002), the China Postdoctoral Science Foundation (Grants Nos. 2021TQ0043 and 2021M700437) and the  National Natural Science Fund for Excellent Young Scientists Fund Program (Overseas).

\bibliography{myRef.bib}

\end{document}